\documentclass[letterpaper,twocolumn,american]{revtex4}
\usepackage[T1]{fontenc}
\usepackage[latin1]{inputenc}
\usepackage{amsmath}
\usepackage{graphicx}
\usepackage{amssymb}

\makeatletter

\providecommand{\LyX}{L\kern-.1667em\lower.25em\hbox{Y}\kern-.125emX\@}

\usepackage{babel}
\makeatother
\begin{document}

\title{Evidence of fractional transport in point-vortex flow}

\author{Xavier Leoncini}

\email{leoncini@cims.nyu.edu}

\affiliation{{\small Courant Institute of Mathematical Sciences, New York University,
251 Mercer St., New York, NY 10012, USA}}

\author{Leonid Kuznetsov}

\email{leonid@cfm.brown.edu}

\affiliation{{\small Lefschetz Center for Dynamical Systems, Division of Applied
Mathematics, Brown University, Providence, RI 02912, USA}}

\author{George M. Zaslavsky}

\email{zaslav@cims.nyu.edu}

\affiliation{{\small Courant Institute of Mathematical Sciences, New York University,
251 Mercer St., New York, NY 10012, USA}}

\affiliation{{\small Department of Physics, New York University, 2-4 Washington
Place, New York, NY 10003, USA }}

\begin{abstract}
Advection properties of passive particles in flows generated by point
vortices are considered. Transport properties are anomalous with characteristic
transport exponent $\mu \sim 1.5$. This behavior is linked back to
the presence of coherent fractal structures within the flow. A fractional
kinetic analysis allows to link the characteristic transport exponent
$\mu $ to the trapping time exponent $\gamma =1+\mu $. The quantitative
agreement is found for different systems of vortices investigated
and a clear signature is obtained of the fractional nature of transport
in these flows.
\end{abstract}
\maketitle

\section{Introduction}

One area extensively studied in the last twenty years is the phenomenon
of chaotic advection \cite{Aref84}-\cite{Crisanti92}. This phenomenon
results from the chaotic nature of Lagrangian trajectories and enhances
the mixing of tracers in laminar flows. Indeed in its absence the
mixing relies on the less efficient molecular diffusion. Hence, its
applications are important in geophysical flows where advected quantities
vary from the ozone in the stratosphere to various pollutants in the
atmosphere and ocean, or such scalar quantities as temperature or
salinity. This growing interest in geophysical flows increases the
relevance of two-dimensional models and more specifically the problem
of advection in a system with many vortices \cite{provenzale99}-\cite{Meleshko93}.
Moreover, different observations and numerous models have shown that
the transport of advected particles is anomalous and can be linked
to the Levy-type processes and their generalizations \cite{Chernikov90}-\cite{Kovalyov2000}.
Another peculiarity of two-dimensional worlds is the presence of the
inverse energy cascade in turbulent flows, which results in the emergence
of coherent vortices, dominating the flow dynamics \cite{Benzi86}-\cite{Carnevale91}. 

In order to tackle this problem and especially the anomalous features,
our approach has been gradual and the present work completes a series
of papers \cite{KZ98}-\cite{LZ02}, which consists of gradual successive
steps of dynamical investigations of transport in two-dimensional
flows. We restricted ourselves to a relatively simple model, and performed
a thorough analysis of the dynamics of tracers. This approach allows
to infer some properties of the kinetics which actually govern transport.
To settle for a model, we recall that systems point vortices have
been used to mimic the dynamics of finite-sized vortices \cite{Zabusky82}-\cite{VFuentes96},
and mention that the evolution of $2D$ turbulence after the emergence
of the vortices has been successfully described by punctuated Hamiltonian
models \cite{Carnevale91,Benzi92,Weiss99}. Moreover, noisy point
vortex dynamics have been recently used to describe exact unstationnary
two-dimensional solution of the Navier-Stockes equation \cite{Agullo01}.
Therefore, despite their relative simplicity, it was clear that systems
of point vortices managed to capture some of the essential features
of two-dimensional flows, it therefore seemed natural to consider
these systems as the basis for our investigations.

In the following we investigate the dynamics as well as the advection
properties of point vortex systems, more specifically we address the
problem with 3 point vortices as well as systems of 4 and 16 vortices.
A system of three point-vortices is integrable and often generate
periodic flows (in co-rotating reference frame)\cite{Aref79}-\cite{Tavantzis88}.
We then can investigate the phase space of passive tracers with Poincar\'e
maps. A well-defined stochastic sea filled with islands of regular
motion is observed, among these are special islands known as {}``vortex
cores'' surrounding each of the vortices. The non-uniformity of the
phase space and the presence of islands of regular motion within the
stochastic sea has a considerable impact on the transport properties
of such systems. The phenomenon of stickiness on the boundaries of
the islands generates strong {}``memory effects'' and transport
is found to be anomalous. On the other hand, the motion of $N$ point
vortices on the plane is generically chaotic for $N\ge 4$ \cite{Novikov78}-\cite{Ziglin80}.
The periodicity is lost when considering a system of four vortices
or more, but snapshots of the system have revealed the cores surrounding
vortices are a robust feature \cite{Babiano94}-\cite{Boatto99},
the actual accessible phase space is in this sense non uniform and
stickiness around these cores has been observed \cite{Laforgia01}.
In fact, a refined analysis has revealed that in systems of 4 vortices
and more the cores are surrounded by coherent jets, within which little
dispersion occurs and thus which give rise to anomalous transport
properties \cite{LZ02}. 

The goal of this paper is to propose a unified view of transport properties
in these system of vortices. In Sec.~\ref{sec:Point-Vortex-and-passive},
we start to recall briefly the general equations governing dynamics
of point vortices as well as passive tracers. We then describe with
more details the motion of vortices and tracers for some specific
cases. We start with integrable systems of vortices namely, a system
with three identical vortices, and a system of with two identical
vortices and one of the opposite sign, a system which allows to set
up parameters in order to set the motion of the vortices on a collapse
course. We then consider a system of four and sixteen identical vortices.
In Sec.~\ref{sec:Transport-properties} we investigate the transport
properties of advected particles in these systems and measure the
typical transport exponent, finally in Sec.~\ref{sec:Fractional-aspects-of}
we develop on the fractional aspects of transport and link its anomalous
features to the fractal nature of the topology of the flow.

\section{Point-Vortex and passive tracer motion\label{sec:Point-Vortex-and-passive}}

\subsection{Basic Equations}

Systems of point vortices are exact solutions of the two-dimensional
Euler equation\begin{eqnarray}
\frac{\partial \Omega }{\partial t}+[\Omega ,\Psi ] & = & 0\label{Euler}\\
\Delta \Psi  & = & \Omega \: ,\label{Poisson}
\end{eqnarray}
where $\Omega $ is the vorticity and $\Psi $ is the stream function.
The vortices describe the dynamics of the singular distribution of
vorticity \begin{equation}
\Omega (z,t)=\sum _{\alpha =1}^{N}k_{\alpha }\delta \left(z-z_{\alpha }(t)\right),\label{vorticity}\end{equation}
 where $z$ locates a position in the complex plane, $z_{\alpha }=x_{\alpha }+iy_{\alpha }$
is the complex coordinate of the vortex $\alpha $, and $k_{\alpha }$
is its strength, in an ideal incompressible two-dimensional fluid.
This system can be described by a Hamiltonian of $N$ interacting
particles (see for instance \cite{Lamb45}), referred to as a system
of $N$ point vortices. The system's evolution writes\begin{equation}
k_{\alpha }\dot{z}_{\alpha }=-2i\frac{\partial H}{\partial \bar{z}_{\alpha }}\: ,\hspace {5mm}\dot{\bar{z}}_{\alpha }=2i\frac{\partial H}{\partial (k_{\alpha }z_{\alpha })}\: ,\label{vortex.eq}\end{equation}
 where the couple $(k_{\alpha }z_{\alpha },\bar{z}_{\alpha })$ are
the conjugate variables of the Hamiltonian $H$. The nature of the
interaction depends on the geometry of the domain occupied by fluid.
For the case of an unbounded plane, the resulting complex velocity
field $v(z,t)$ at position $z$ and time $t$ is given by the sum
of the individual vortex contributions:\begin{equation}
v(z,t)=\frac{1}{2\pi i}\sum _{\alpha =1}^{N}k_{\alpha }\frac{1}{\bar{z}-\bar{z}_{\alpha }(t)}\: ,\label{velocity}\end{equation}
 and the Hamiltonian becomes\begin{equation}
H=-\frac{1}{2\pi }\sum _{\alpha >\beta }k_{\alpha }k_{\beta }\ln |z_{\alpha }-z_{\beta }|\label{Hamiltonianvortex}\end{equation}
 The translational and rotational invariance of the Hamiltonian $H$
provides for the motion equations (\ref{vortex.eq}) three other conserved
quantities, besides the energy, \begin{equation}
Q+iP=\sum _{\alpha =1}^{N}k_{\alpha }z_{\alpha },\hspace {1.2cm}L^{2}=\sum _{\alpha =1}^{N}k_{\alpha }|z_{\alpha }|^{2}.\label{constantofmotion1}\end{equation}
Among the different integrals of motion, there are three independent
first integrals in involution: $H$, $Q^{2}+P^{2}$ and $L^{2}$;
consequently the motion of three vortices on the infinite plane is
always integrable and chaos arises when $N\ge 4$ \cite{Novikov75}.

The evolution of a passive tracer is given by the advection equation\begin{equation}
\dot{z}=v(z,t)\label{gen.adv}\end{equation}
 where $z(t)$ represent the position of the tracer at time $t$,
and $v(z,t)$ is the velocity field (\ref{velocity}). For a point
vortex system, the velocity field is given by Eq. (\ref{velocity}),
and equation (\ref{gen.adv}) can be rewritten in a Hamiltonian form:
\begin{equation}
\dot{z}=-2i\frac{\partial \Psi }{\partial \bar{z}},\hspace {5mm}\dot{\bar{z}}=2i\frac{\partial \Psi }{\partial z}\end{equation}
 where the stream function \begin{equation}
\Psi (z,\bar{z},t)=-\frac{1}{2\pi }\sum _{\alpha =1}^{4}k_{\alpha }\ln |z-z_{\alpha }(t)|\label{stream}\end{equation}
 acts as a Hamiltonian. The stream function depends on time through
the vortex coordinates $z_{\alpha }(t)$, implying a non-autonomous
system.

Due to chaotic nature of the evolutions we rely heavily on numerical
simulations. The trajectories of the vortices as well as those of
the passive tracers are integrated numerically using the symplectic
scheme described in \cite{McLachlan92} and which has already been
successfully used in \cite{KZ98,KZ2000,LKZ2000,LKZ01,Laforgia01,LZ02}.

\subsection{3-Vortex systems}

A general classification of different types of three vortex motion,
as well as studies of special cases were addressed by many authors
\cite{Synge49,Novikov75,Aref79,Tavantzis88}. Among the different
possibilities quasiperiodic motion of the vortices is found generically
for solutions for which the motion of the vortices is bound within
a finite domain. We consider two different type systems. On one hand
we consider a system with three identical vortices (see \cite{KZ98}),
in this setting the two extreme regimes for the advection pattern
of strong and weak chaos can be investigated with good control and
analytical expression of vortex core is given. On the other hand in
order to stress our results we consider also 3-vortex systems in the
vicinity of a configuration leading to finite time singular solution.
Indeed three vortices can be brought to a single point in finite time
by their mutual interaction. Aref \cite{Aref83} points out that this
result was already known to Gr\"{o}bli more than a century ago. This
phenomenon, known as \textit{point-vortex collapse} was studied in
\cite{Aref79,Novikov79,Tavantzis88,Kimura90}. Thus, under certain
conditions, which depends both on the initial conditions, and the
vortex strengths, the motion is self similar, leading either to the
collapse of the three vortices in a finite time, or by time reversal,
to an infinite expansion of the triangle formed by the vortices. Let
us re-derive these conditions of a collapse or an infinite expansion.
For a system of three vortices in an unbounded domain, the invariance
of the Hamiltonian (\ref{Hamiltonianvortex}) under translations allows
a free choice of the coordinate origin, which we put to the center
of vorticity (when it exists), the angular momentum in a frame independent
form is then rewritten as:\begin{eqnarray}
K & = & \left[\left(\sum _{i}k_{i}\right)L^{2}-P^{2}\right]\label{eq:K}\\
 & = & \frac{1}{2}\sum _{i\ne j}k_{i}k_{j}|z_{i}-z_{j}|^{2}
\end{eqnarray}
The first condition is immediate as for a collapse to occur the frame
independent angular momentum (\ref{eq:K}) has to vanish. For the
second condition we look for conditions resulting in a scale invariant
Hamiltonian. With such spirit let us divide all lengths by a common
factor $\lambda $ in the Hamiltonian $H$. We readily obtain $H'(\lambda )=H+\left(\sum k_{i}k_{j}\right)\ln \lambda $,
we then obtain the collapse conditions:\begin{equation}
K=0\: ,\hspace {10mm}\sum _{i}\frac{1}{k_{i}}=0\: ,\label{eq:Collapse-condition}\end{equation}
 \emph{i.e}, the total angular momentum in its frame free form and
the harmonic mean of the vortex strengths are both zero (\ref{eq:Collapse-condition}).
Near a collapse configuration, the two conditions (\ref{eq:Collapse-condition})
allow two different ways to approach the singularity. Namely, we can
change initial conditions which changes the value of $K$, or change
the vortex strength and modify the harmonic mean. These lead to different
types of motion which have been classified and studied in \cite{LKZ2000}.

In all considered cases, the relative vortex motion is periodic, i.e.
the vortex triangle repeats its shape after a time $T$. This does
not imply a periodicity of the absolute motion, since the triangle
is rotated by some angle $\Theta $ during this time. In general,
$\Theta $ is incommensurate with $2\pi $, rendering a quasiperiodic
time dependence of $z_{l}(t)$. Let us consider a reference frame,
rotating around the center of vorticity with an angular velocity $\Omega \equiv \Theta /T.$
In this co-rotating reference frame, vortices return to their original
positions in one period of relative motion $T$, their new coordinates
\[
\tilde{z}\equiv z\: e^{-i\Omega t}\: ,\]
 are periodic functions of time. In the co-rotating frame the advection
equation retains its Hamiltonian form with a new stream function $\tilde{\Psi }$.
In this frame $\tilde{\Psi }$ is time-periodic ($\tilde{\Psi }(\tilde{z},\tilde{\bar{z}},t+T)=\tilde{\Psi }(\tilde{z},\tilde{\bar{z}},t)$)
and well-developed techniques for periodically forced Hamiltonian
systems can be used to study its solutions \cite{KZ98}. Note, that
the one-period rotation angle $\Theta $ is defined modulo $2\pi $,
making the choice of the co-rotating frame non-unique.

\begin{figure}[htbp]
\begin{center}\includegraphics[  width=7cm]{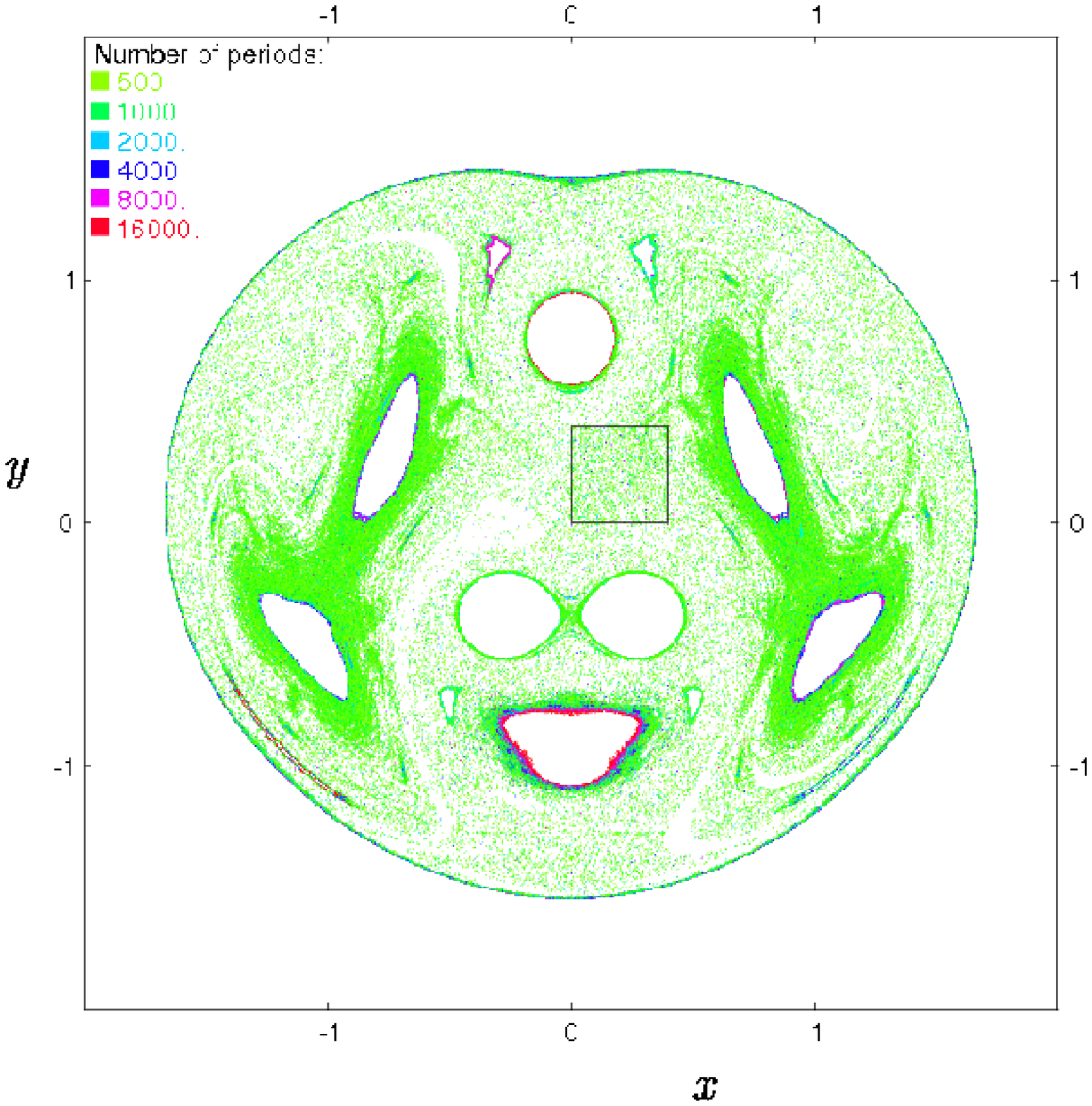}\end{center}

\begin{center}\includegraphics[  width=7cm]{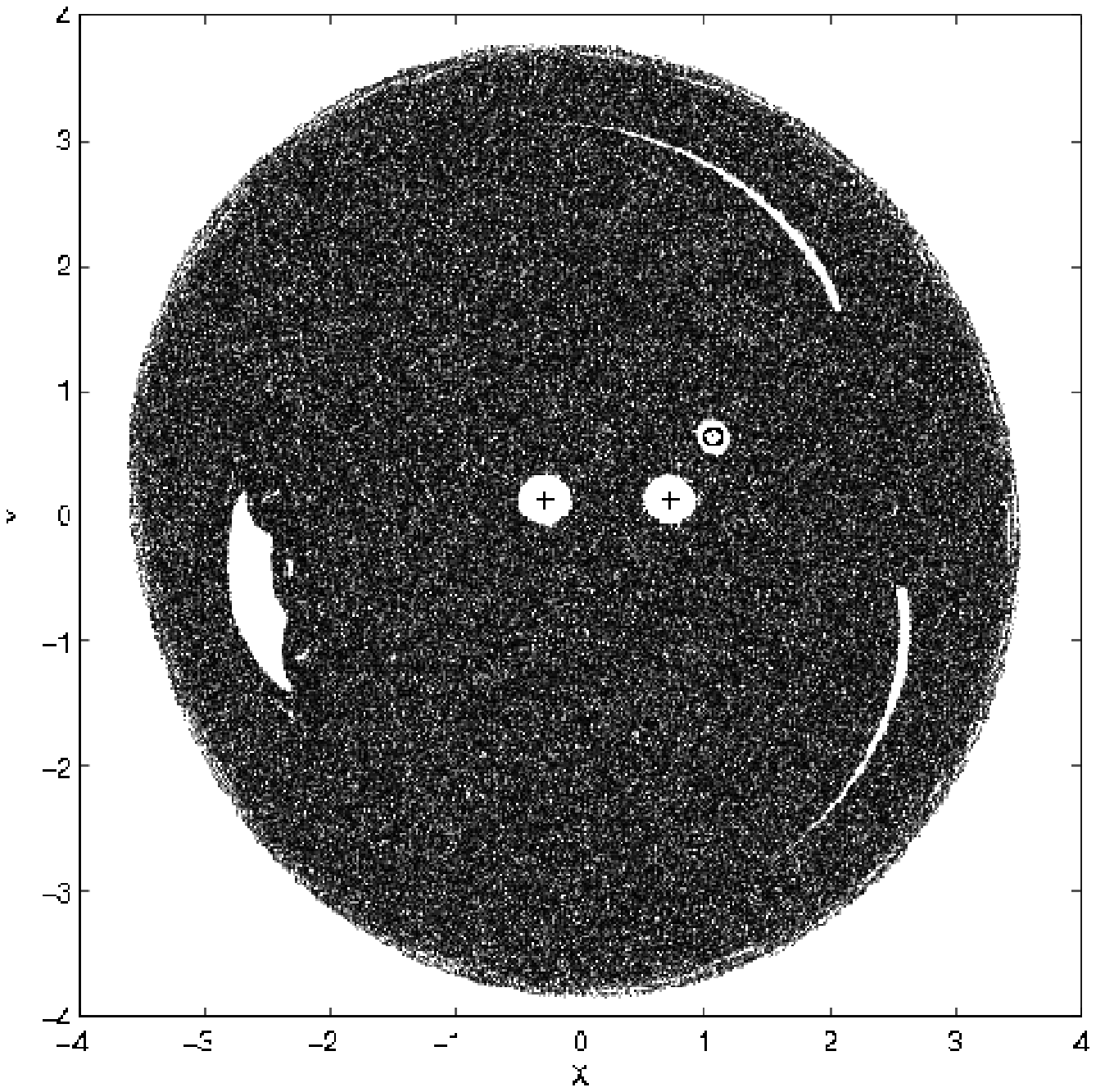}\end{center}

\caption{Poincar\'{e} Map for strongly chaotic system (upper plot) and the
system in the close to collapse configuration (lower plot). \label{cap:Poincar-Map}}
\end{figure}

Anomalous properties of tracer advection in a flow generated by the
motion of three identical vortices were analyzed in \cite{KZ2000}.
The vortices were initialized with such initial conditions creating
a large region of chaotic tracer motion. The structure of the chaotic
region is quite complex, with an infinite number of KAM-islands (stratified
with regular trajectories) of different shapes and sizes embedded
into it. To visualize these structures, we construct Poincar\'{e}
sections of tracer trajectories (in the co-rotating frame). A Poincar\'{e}
section is defined as an orbit of a period-one (Poincar\'{e}) map
$\hat{P}$\begin{equation}
\hat{P}z_{0}=\tilde{z}(T,z_{0})=e^{-i\Theta }z(T,z_{0})\end{equation}
 where $\tilde{z}(t,z_{0})$ denotes a solution $\tilde{z}(t)$ with
an initial condition $\tilde{z}(t=0)=z_{0}$. Plots of Poincar\'{e}
sections for three identical vortices in the strong chaotic regime
as well as for a system near a collapse configuration are shown in
Figure~\ref{cap:Poincar-Map}%
. We recall that vortex and tracer trajectories were computed using
a symplectic Gauss-Legendre scheme \cite{McLachlan92}. The exact
conservation of Poincar\'{e} invariants by the symplectic scheme
suppresses numerical diffusion, yielding high-resolution phase space
portraits.

The Poincar\'{e} sections are presented in the Figure~\ref{cap:Poincar-Map}
show an intricate mixture of regions with chaotic and regular tracer
dynamics, typical for periodically forced Hamiltonian systems. All
three phase portraits share common features with the advection patterns:
the stochastic sea is bounded by a more or less circular domain, there
are a islands inside it, where the tracer's motion is predominantly
regular. In particular, all three vortices are surrounded by robust
near-circular islands, known as vortex cores. An expression of the
radius of the cores is computed when the vortices are equal \cite{KZ98},
and the minimum inter-vortex distance through time provides as well
a good upper estimate of the core radii (see for instance \cite{Babiano94,LKZ01}),
this estimate proves also to be quantitative in a four vortex system
\cite{Laforgia01} as well as in 16-vortex systems\cite{LZ02}.

\subsection{4 and 16-vortex systems}

Due to the generic chaotic nature of 4-point vortex system, understanding
the vortex motion necessitates a different approach than for integrable
3-vortex systems. For identical vortices it is possible to perform
a canonical transformation of the vortex coordinates \cite{ArefPomp82}.
For a 4-vortex system, this transformation results in an effective
system with $2$ degrees of freedom, providing a conceptually easier
framework in which perform well defined two-dimensional Poincar\'e
sections can be performed \cite{ArefPomp82,Laforgia01}. To summarize
the results obtained in \cite{ArefPomp82,Laforgia01}, the motion
is in general chaotic, except for some special initial conditions,
for instance when the vortices are forming a square the motion is
periodic and the vortices rotate on a circle, then symmetric deformation
($z_{3}=-z_{1}$ and $z_{4}=-z_{2}$) of the square lead to quasiperiodic
motion (periodic motion in a given rotating frame).

As a prerequisite to our investigations on the advection of passive
tracers, a basic understanding of the vortex subsystem behavior is
necessary. For this matter, an arbitrary initial condition of the
4-vortex system is chosen, although the Poincar\'e section is computed
and the desired generic chaotic behavior verified \cite{Laforgia01}.
As we evolve from the 4 vortex system to 16 vortices the phase space
dimension is considerably increased and due to the long range interaction
between vortices (see the Hamiltonian \ref{Hamiltonianvortex}) the
energy does not behave as an extensive variable. Thus, in order to
keep some coherence between the four vortex system and the sixteen
vortex one, we chose to keep the average area occupied by each vortex
approximatively constant. The switch from 4 to 16 vortices can then
be thought of as increasing the domain with non-zero vorticity while
keeping the vorticity {}``uniform'' within the patch. The initial
condition is chosen randomly within a disk and there is no vortices
with close neighborhood to avoid any possible forced pairing. After
that all position are rescaled to match the condition of uniform vorticity. 

We shall now consider some specific behavior of these chaotic dynamics,
namely the phenomenon of vortex pairing. The simulations performed
in \cite{Laforgia01,LZ02} indicate that long time vortex pairing
exists, in fact the formation of long-lived triplet (a system of $3$
bound vortices) is also observed \cite{LZ02}, and thus the relevance
of three vortex systems is confirmed. In fact, the formation of triplets
or pair of vortices concentrates vorticity in small regions of the
plane and in some sense is reminiscent of what is observed in 2D turbulence.
However, since no quadruples (or even larger clusters) are observed,
we may speculate that this concentration of vorticity is intimately
linked to the dynamics and long memory effects associated with stickiness.
Indeed this phenomenon can be thought as sticking phenomenon to an
object of lesser dimension than the actual phase space with some constraints:
the object is reached by generating subsystems (2 and 3 vortex systems)
whose integrability is a good approximate for a fairly long time.
For comparison we mention that a typical time of an eddy turnover
used in \cite{Weiss98}, corresponds to a time of order $\Delta t\sim 1-5$
in these systems while triplets and pairing are typical times are
at least 2 orders of magnitude larger.

We have now a sufficient knowledge of the dynamics of the chaotic
vortex systems, and move on to the behavior of passive tracers generated
by these flows. %
\begin{figure}[htbp]
\begin{center}\includegraphics[  width=7cm]{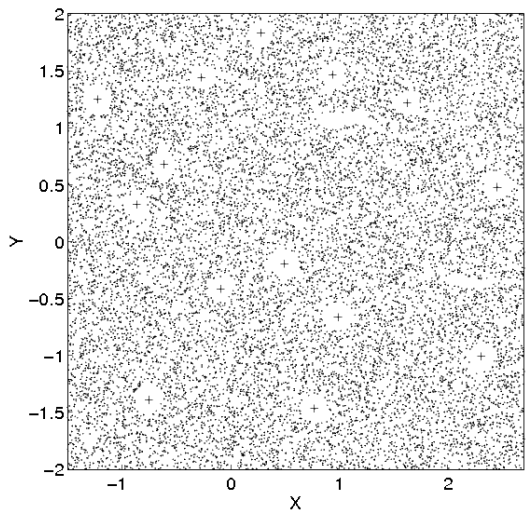}\end{center}

\caption{Local snapshot of the system of $16$ vortices with $9.\: 10^{4}$
tracers. The vortices are located with the {}``+'' sign. We can
see the cores surrounding the vortices.\label{Figsnapshot16v}}
\end{figure}
 For the system of 4 point vortices, successive snapshots have shown
that passive tracers can stick on the boundaries of cores and jump
from one core to another core during a pairing or escape from the
core due to perturbations %
\begin{figure}[htbp]
\begin{center}\includegraphics[  width=7cm,
  keepaspectratio]{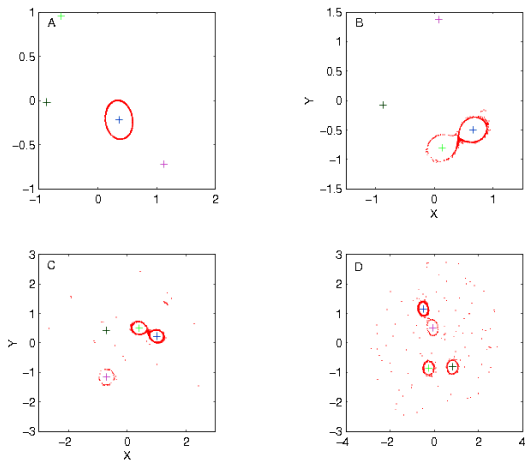}\end{center}

\caption{Four consecutive snapshots for the four vortex system and 1000 particles.
These correspond to four consecutive pairing. Tracers are initially
placed around one vortex. As pairings occur, some jump from a vortex
core to another where they remain after. While the vortex-pairing
occurs some particles also escape from the cores. We notice that after
four pairings all cores have been {}``contaminated'' and are populated
with tracers originating from the first core, while about 10\% of
tracers have escaped from the sticky region surrounding the cores.
\label{coreexchange}}
\end{figure}
. The fact that a tracer is able to escape from a core means that
the surrounding regions of the cores are connected to the region of
strong chaos. The results presented in \cite{Laforgia01} indicate
that these regions mix poorly with the region of strong chaos. One
way to track this phenomenon is to use Finite-Time Lyapunov exponents
(FTLE) and to eliminate domains of small values of the FTLE \cite{Boatto99,Castiglione99,Andersen00}.
Once these exponents are measured from tracers' trajectories whose
initial conditions are covering the plane, a scalar field distributed
within the space of initial conditions is obtained and the two dimensional
plot of the scalar field reveals regions of vanishing FTLE, namely
the cores surrounding the vortices and the far field region. The cores
are thus regions of small FTLE, meaning that two nearby trajectories
are bound together for long times, and this despite the core's chaotic
motion. These properties reveal typically a sharp change of the tracers
dynamics as it crosses from the region of strong chaos to the core.

\section{Transport properties\label{sec:Transport-properties}}

\subsection{Definitions }

In general the deterministic description of the motion of a passive
particle in the chaotic region is impossible, a local instability
produces exponential divergence of trajectories. Even the outcome
of an idealized numerical experiment is non-deterministic, indeed
in this situation round-off errors are creeping slowly but steadily
from the smallest to the observable scale. The long-time behavior
of tracer trajectories in the mixing region is therefore studied using
a probabilistic approach. In the absence of long-term correlations,
a kinetic description, which uses the Fokker-Plank-Kolmogorov equation
and leads to Gaussian statistics, \cite{Zaslavsky92} works fairly
well for many situations. Yet in the present case, the topology of
the advection pattern (Fig.~\ref{cap:Poincar-Map}) and the trapping
of tracers in the neighborhood of cores (see Fig.~\ref{coreexchange})
indicate that anomalous statistical properties of the tracers should
be expected. The singular zones around KAM islands and the cores give
often rise to a stickiness phenomenon and produce long-time correlations,
which result in profound changes in the kinetics. In some cases these
{}``memory effects'' can be accounted for by the modification of
the diffusion coefficient in the FPK equation \cite{Chirikov79,Rechester80},
but often their influence is more profound \cite{Zaslavsky93,dCN98,Castiglione99,KZ2000,LKZ01,LZ02},
and leads to a super-diffusive behavior with faster than linear growth
of the particle displacement variance: \begin{equation}
\langle (x-\langle x\rangle )^{2}\rangle \sim t^{\mu }\: ,\label{anom}\end{equation}
the transport exponent $\mu $ exceeds the Gaussian value: $\mu >1$.

For all considered case, the vortices are moving within a finite domain.
It is thus important to define what quantities will be measured to
characterize the transport properties of the system. There has been
evidence in \cite{Boatto99} of radial diffusion, but the diffusion
coefficient $D$ is vanishing as $R\rightarrow \infty $ with typical
behavior $D\sim 1/R^{6}$. In the case of more than four vortices
we can still expect a similar type of behavior. However the region
far from the vortices is of little interest when one want to address
the transport properties of typical geophysical flows and the most
relevant is the region being accessible to the vortices also called
the region of {}``strong chaos''. In the 3-vortex systems the question
is even more crucial as the accessible domain of tracers (the chaotic
sea) reduces to a finite region surrounded by a KAM curve. One way
around this finite domain problem is to focus our interest on the
character of tracer rotation, and for that matter, we define its azimuthal
coordinate in the center of vorticity reference frame \begin{equation}
\theta (t)\equiv {\mbox {Arg}}\, \, z\end{equation}
 to be a continuous function of time, i.e. $\theta (t)\in (-\infty ,\infty )$
keeps track of the number of revolutions performed by a tracer. However
for many vortices we choose another quantity, namely we consider tracer
transport by measuring the arclength $s(t)$ of the path traveled
by an individual tracer up to a time $t$. The arclength $s(t)$ writes\begin{equation}
s_{i}(t)=\int _{0}^{t}v_{i}(t')dt'\: ,\label{arclenghtdefinition}\end{equation}
where $v_{i}(t')$ is the absolute speed of the particle $i$ at time
$t'$. The main advantage of this quantity is that it is independent
of the coordinate system and as such we can expect to infer intrinsic
properties of the dynamics. The main observable characteristics will
be moments of the angle $\theta (t)$ or distance $s(t)$:\begin{equation}
M_{q}=\langle |x_{i}(t)-\langle x_{i}(t)\rangle |^{q}\rangle \: ,\label{momentsdefi}\end{equation}
where $i$ corresponds either to the i-th vortex or a tracer in the
field of 3, 4 or 16 vortices and $x$ stands for $s$ or $\theta $.
The averaging operator $\langle \cdots \rangle $ needs a special
comment. Expecting anomalous transport one should be ready to have
infinite moments starting from $q\ge q_{0}$. To avoid any difficulty
with infinite moments we consider truncated distribution function,
which was discussed in details in \cite{Weitzner01} and allows to
satisfy the physical restriction of a finite velocity and and the
finite time of our simulations. This condition actually put some constraints
on the maximum meaningful value $q*$ of $q$ , and beyond $q*$ the
moments are basically monitoring the population of almost ballistic
trajectories.

Up to the mentioned constraints, we will always consider the operation
of averaging to be performed over truncated distributions. In this
perspective all moments are finite and one can expect\begin{equation}
M_{q}\sim D_{q}t^{\mu (q)}\: ,\label{momentexpectation}\end{equation}
with, generally, $\mu (q)\ne q/2$ as is expected from normal diffusion.
The nonlinear dependence of $\mu (q)$ is a signature of the multifractality
of the transport, . For more information on the appearance of multi-fractal
kinetics and related transport see \cite{Zaslav2000,Zaslav2000_1}.
Some authors use the notion of weak ($\mu (q)=const\cdot q$) and
strong ($\mu (q)\ne const\cdot q$) anomalous diffusion \cite{Castiglione99,Andersen00}
or strong and weak self-similarity \cite{Ferrari01}.

Recently \cite{Castiglione99,Andersen00}, two types of anomalous
diffusion were distinguished by the behavior of the moments, and the
notion of weak and strong anomalous diffusion \cite{Castiglione99,Andersen00}
or identically strong and weak self-similarity \cite{Ferrari01} introduced.
When the evolution of all of the moments can be described by a single
self-similarity exponent $\nu $ according to \begin{equation}
\mu (q)=\nu \cdot q\label{eq:weak}\end{equation}
 refers to {}``weak anomalous diffusion'', whereas the case when
$\nu $ in (\ref{eq:weak}) is not constant, i.e. \begin{equation}
\mu (q)=q\nu (q)\; ,\hspace {5mm}\nu (q)\ne const\label{eq:strong}\end{equation}
 is named {}``strong anomalous diffusion''. This distinction is
important since in the weak case the PDF must evolve in a self-similar
way: \begin{equation}
P(x,t)=t^{-\nu }f(\xi ),\hspace {1cm}\xi \equiv t^{-\nu }(x-\langle x\rangle )\label{selfsim}\end{equation}
 whereas non-constant $\nu (q)$ in (\ref{eq:strong}) precludes such
self-similarity (see also the discussions in \cite{KZ2000,LKZ01}
for details about the non self-similar behavior).

\begin{figure}[htbp]
\begin{center}\includegraphics[  width=7cm]{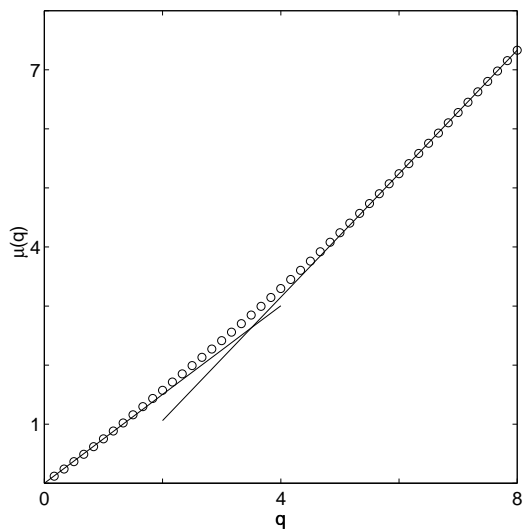}\end{center}

\caption{The exponent $\mu (q)$ versus the moment order $q$ for the angle
distribution ($\langle |\theta (t)-\overline{\theta }(t)|^{q}\rangle \sim t^{\mu (q)}$)
is plotted for a 3-point vortex system ($t>1.5\; 10^{5}$). We notice
two linear behaviors: $\left\{ \protect\begin{array}{llll}
 q<2 & , & \mu (q)= & 0.75q\protect\\
 q>2 & , & \mu (q)= & 1.04q-Cte\protect\end{array}
\right.$. Vortex strengths are $(-0.3,\: 1,\: 1)$. The period of the motion
is $T=17.53$.\label{cap:The-exponent}}
\end{figure}

In all our simulations, the results show that the transport of tracers
in point-vortex systems is strongly anomalous and super-diffusive,
hence to avoid redundancy we graphically only present the transport
properties of passive tracers obtained for a 3-point vortex flow in
Fig.~\ref{cap:The-exponent}. The behavior for the other considered
systems is very similar, namely our results show, that for all considered
systems $\mu (q)$ is well approximated by a piecewise linear function
of the form: \begin{equation}
\mu (q)=\left\{ \begin{array}{ccc}
 \nu q & \mbox {for} & q<q_{c}\\
 q-c & \mbox {if} & q>q_{c}\end{array}
\right.\label{Cform}\end{equation}
 where $c$ is a constant, and $q*$ is the crossover moment number.

Even though we only considered few different initial condition for
the different vortex system, it is reasonable to assume that the transport
properties obtained for such systems are fairly general since all
give the same kind anomalous behavior with a transport exponent more
or less around $\mu (2)\pm 1.5$. %

\section{Fractional aspects of transport\label{sec:Fractional-aspects-of}}

\subsection{Poincar\'{e} Recurrences and trapping times exponent}

The origin of the anomalous transport properties can be linked back
to the intermittent character of tracer motion. This phenomenon is
characterized by an anomalous distribution of recurrences of the Poincar\'{e}
map of tracer trajectories for the three vortex systems, or the algebraic
decay of the density of trapping time within jets in many vortex systems.
To define recurrences, we take a region $B$ in the chaotic sea, and
register all returns of a Poincar\'{e} map trajectory into $B$.
The length of a recurrence is a time interval between two successive
returns. The distribution of the recurrence times does not depend
on the choice of the trajectory in the chaotic region, in dynamical
systems with perfect mixing this distribution is Poissonian but the
Hamiltonian systems with coexisting regular and chaotic motions exhibit
a power-law tails in the distribution. Recurrence distributions for
tracers in three vortex systems show that all distributions have long
tails, indicating, that between the returns tracers are being trapped
in long flights of highly correlated motion. Long recurrences are
distributed according to a power law \begin{equation}
P(\tau )\sim \tau ^{-\gamma }\end{equation}
 The measured values of the exponent $\gamma $ in Refs.\cite{KZ2000,LKZ01}
are all around: \begin{equation}
\gamma \sim 2.5.\end{equation}

The value of the decay exponent $\gamma $ does not depend on the
choice of the domain $B$ as long as it is taken in the well-mixed
region, away from the KAM-islands.

\begin{figure}[htbp]
\begin{center}\includegraphics[  width=7cm,
  keepaspectratio]{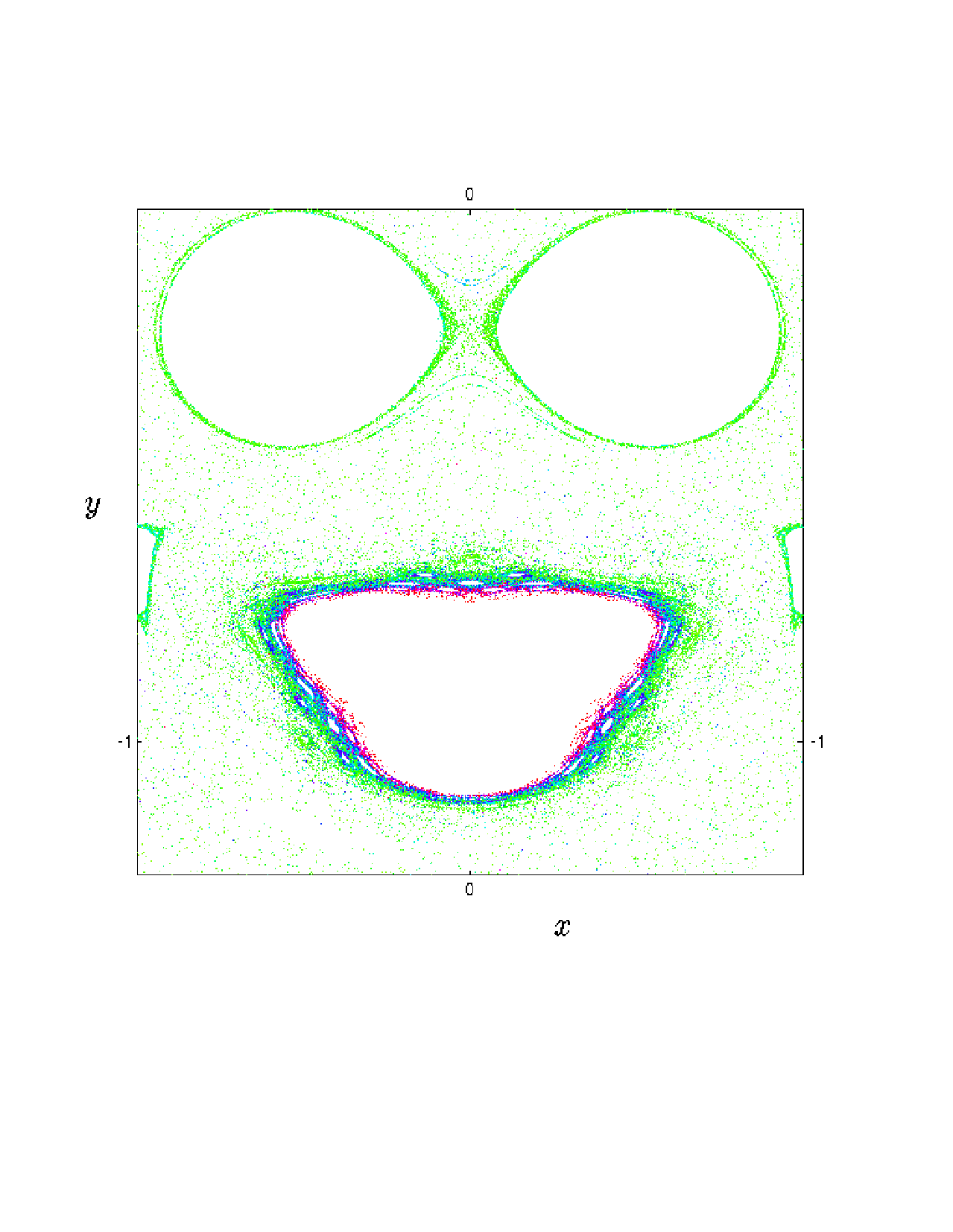}\end{center}

\begin{center}\includegraphics[  width=7cm,
  keepaspectratio]{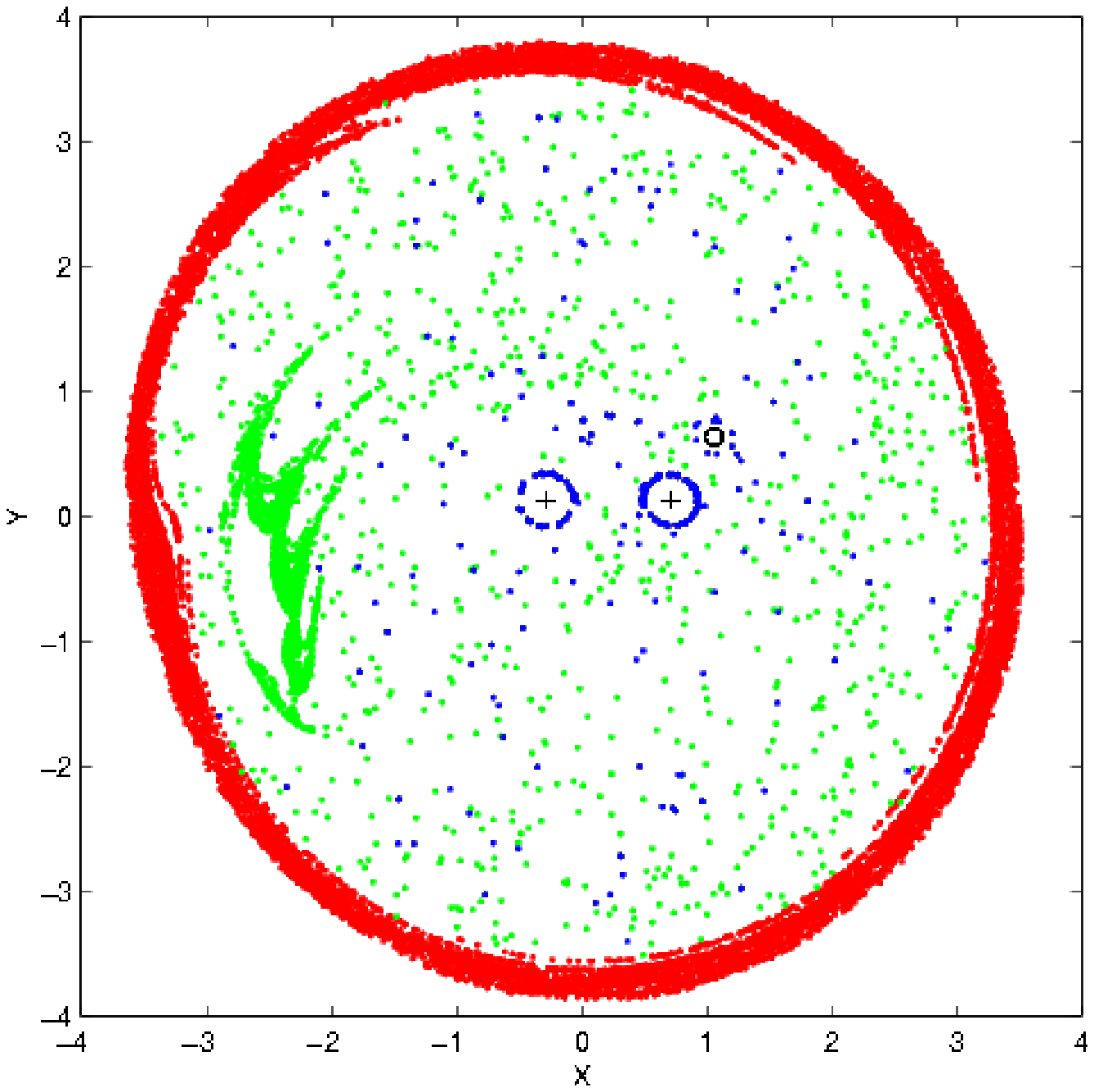}\end{center}

\caption{Visualization of the fractal origin of anomalous transport in three
vortex flows. The particles whose trajectories have longest return
times penetrate deeply into the hierarchical island-around-island
structures (upper plot). Stickiness to islands corresponding to different
type of characteristic average speed reveals the multi-fractal nature
of transport (lower plot).\label{cap:3vortfrac}}
\end{figure}

In order to establish the origin of the long recurrences the corresponding
Poincare cycles (i.e. parts of the map orbit between the returns)
are plotted color-coded by the recurrence times $\tau $, see Fig.~\ref{cap:3vortfrac}
(only the cycles with $\tau >10^{4}\gg \langle \tau \rangle \approx 850$
are shown). The orbits of the long cycles are concentrated at the
boundaries of the chaotic sea, around the islands of regular motion
and at the external boundary. The longest recurrences (shown in red)
correspond to the trajectories that penetrate deeply into the hierarchical
island-around-island structures, see Fig.~\ref{cap:3vortfrac}. This
phenomenon, known as stickiness (of KAM island boundaries) introduces
long correlations to the tracer motion and leads to anomalous kinetic
properties of chaotic trajectories. Another way to characterize this
stickiness phenomenon using distribution of time averaged speed has
been used in \cite{LKZ01}, typically each sticky zone corresponds
to a special time averaged speed different than the whole phase space
average, hence sticky portions of particles trajectories can be identified
(see Fig.~\ref{cap:3vortfrac}). %
\begin{figure}[htbp]
\begin{center}\includegraphics[  width=7cm,
  keepaspectratio]{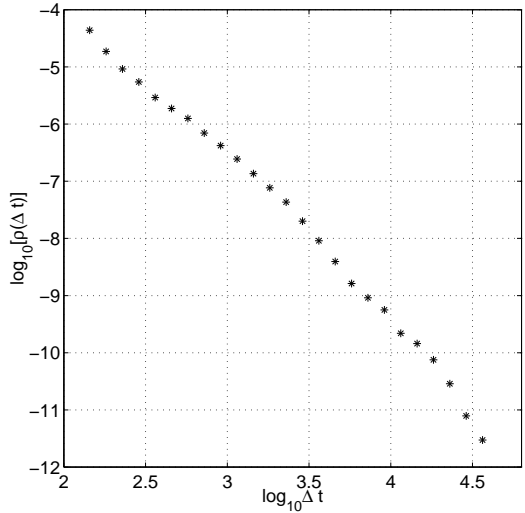}\end{center}

\caption{Tail of the distribution of trapping time intervals $\Delta t$ .
We notice a power-law decay, with some oscillations. Typical exponent
is $\rho (t)\sim t^{-\gamma }$ with $\gamma \approx 2.823$.\label{Figescapetime4vortex}}
\end{figure}

When dealing with more than three vortices the chaotic nature of the
vortex systems does not allow to use the Poincar\'{e} map. However
we can circumvent this problem by actually noticing that when dealing
with more practical situations, we are facing a {}``coarse grained''
phase space, and each point is actually a ball from which infinitely
many real trajectories can depart. Given this fact we can imagine
that two nearby real trajectories diverge exponentially for a while
but then get closer again without actually going to far from each
other, a process which may happen over and over in the case of stickiness
observed around islands in three vortex systems. From the {}``coarse
grained'' perspective those two real trajectories are identical.
We then can infer that there exists bunch of nearby trajectories which
may remain within a given neighborhood for a given time, giving rise
to what is commonly called a \emph{jet} \cite{LZ02,Afanasiev91}.
Note that the stickiness to a randomly moving and not well determined
in phase space coherent structure imposes existence of jets, while
the opposite may not be the case.

To actually measure the jets properties of the system, we use the
following strategy. Let us consider a given trajectory $\mathbf{r}(t)$
evolving within the phase space. For each instant $t$, we consider
a ball $B(\mathbf{r}(t),\delta )$ of radius $\delta $ centered on
our reference trajectory. We then start a number of trajectories within
the ball at a given time, and measure the time it actually takes them
to escape the ball, and have then access to trapping time distribution
which is plotted for a system driven by four vortices in Fig. \ref{Figescapetime4vortex}
and log-log plot clearly shows the power-law decay of the trapping
times, with typical exponent $\gamma =2.82$.

\subsection{Jets and stickiness}

Given the trapping time distribution it is possible to compute a distribution
of finite-time Lyapunov exponent (see \cite{LZ02}), which identifies
a threshold $\sigma _{D}*$ and allows to dynamically detect a jet.
Two different types of jets are identified, namely slow jets evolving
in the region far from the vortices as well as fast jets localized
on the boundaries of vortex cores. These last jets are profoundly
influenced by the phenomenon of vortex pairing, and this phenomenon
is at the origin of some mixing as it actually leads to the division
and merging jets (see Fig. \ref{coreexchange}). 

This localization of the jets confirms the results already illustrated
in Fig. \ref{coreexchange} that the boundaries of the core exhibit
the stickiness. However we insist on the fact that the stickiness
of the system has been confirmed by looking for jets. In this sense
the method used is rather general, and could be applied to other Hamiltonian
systems.

\begin{figure}[htbp]
\begin{center}\includegraphics[  width=7cm,
  keepaspectratio]{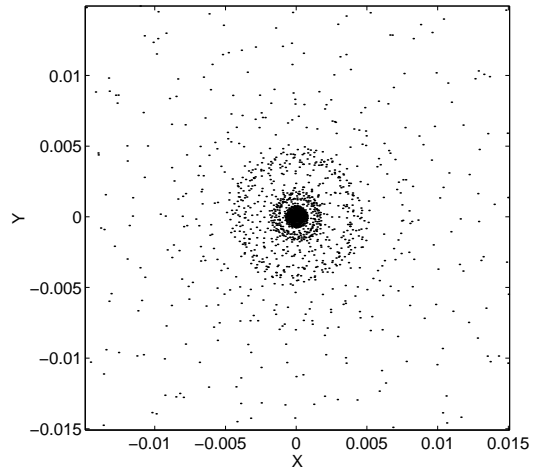}\end{center}

\caption{Relative evolution of a tracer with respect to another within a long
lived jets located in the far field region of the flow generated by
four vortices. The distribution within the jets is gives rise to an
hidden order organized as {}``matroshkas'' (a nested set of jets
with increasing radii). \label{Figslowjet}}
\end{figure}
 We now focus on the inner structure of the jet, a first plot of the
evolution of a relative position of a tracer with respect to another
tracer's position while both are within a jet. A plot of such structure
is presented in Fig. \ref{Figslowjet}, where we discover a fine structure
formed by a hierarchy of circular (tubular) jets within jets. %
\begin{figure}[htbp]
\begin{center}\includegraphics[  width=7cm,
  keepaspectratio]{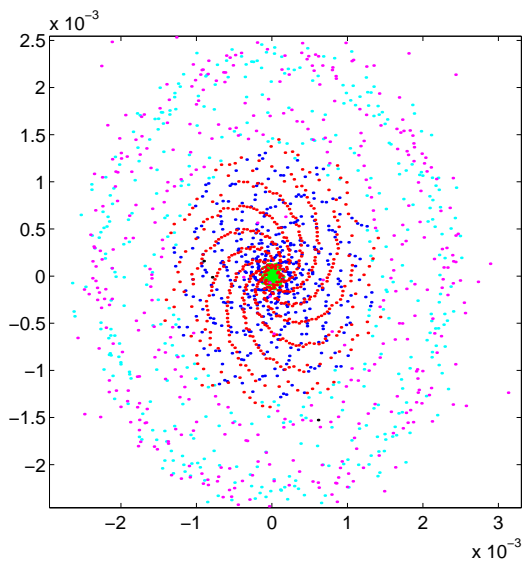}\end{center}

\caption{Jet structure for a long lived jet located in the region of strong
chaos for a system of $16$ vortices. The colors are characterizing
different moments of the life of the jet corresponding to approximatively
equal time intervals. They chronologically range as cyan<blue<green<red<magenta.
We see a similar structure of jets within jets as observed in Fig.
\ref{Figslowjet}, and the tracer spiraling back and forth between
them.\label{Fig16vjettime}}
\end{figure}
 This last structure found for a 4-vortex system, we also looked at
a jet in a system of sixteen vortices. The results are plotted in
Fig. \ref{Fig16vjettime}, where the relative position of the ghost
(relative tracer) is colored differently for different time periods
of the life of the jet. We can see that effectively the nested set
of jets within jets remains, and that the ghosts is also spiraling
back and forth in between. We also see the ghost going back very close
to the tracer. This hierarchical structure is also reminiscent of
the discrete renormalization group, and we speculate that log-periodic
oscillation described in \cite{Benkadda99} should be observed.

\subsection{Kinetics}

In some of the previous publications (see, for example, \cite{Chernikov90,ZEN97,Zaslavsky2000,Zasalvskypreprint})
it was clearly indicated that the properties of anomalous transport
are sensitive to phase space topology. More specifically, if we use
the fractional kinetic equation \cite{Zaslavsky92,ZEN97} in the form
\begin{equation}
{\frac{\partial ^{\beta }P(\theta ,t)}{\partial t^{\beta }}}=\mathcal{D}{\frac{\partial ^{\alpha }P(\theta ,t)}{\partial |\theta |^{\alpha }}}\label{eq:5.1}\end{equation}
 to describe distributions $P(\theta ,t)$ of rotations over angle
$\theta $, then the transport coefficient $\mathcal{D}$ and exponents
$(\alpha ,\beta )$ depend on the presence of different structures
such as boundaries of the domain, islands, cantori, etc.

The caption \ref{cap:3vortfrac} shows that stickiness of trajectories
to specific structures occurs with a filamentation of sticky domains
along stable/unstable manifolds. In fact, different sticky domains
generate different intermittent scenarios with some associated values
of $(\alpha ,\beta )$ \cite{ZEN97,Zaslavsky2000}. As a result, the
real kinetics is multi-fractional and can be characterized by a set
of values of $(\alpha ,\beta )$ or, more precisely, by a spectral
function of $(\alpha ,\beta )$ in the same sense as the spectral
function for multi-fractals \cite{Hentschel84,Frish85,Jensen85}.
The Figure \ref{cap:3vortfrac} characterizes the fact that trajectories,
sticking to different structures (islands), have different angular
velocities. Due to this, different asymptotic to the distribution
function $P(\theta ,t)$ and different values of $(\alpha ,\beta )$
will appear for different time intervals. In other words, for a considered
time interval one can expect a specific {}``intermediate asymptotic''
for $P(\theta ,t)$ and, correspondingly, different pairs $(\alpha ,\beta )$.
Different classes of universality for the values $(\alpha ,\beta )$
were discussed in \cite{Zaslavsky2000}. We will now remind the consequences
of some of these results. 

Multiplying (\ref{eq:5.1}) by $|\theta |^{\alpha }$ and integrating
it over $|\theta |$ we obtain \begin{equation}
\langle |\theta |^{\alpha }\rangle \sim t^{\beta }\label{equ5.2}\end{equation}
 or, in the case of self-similarity the transport exponent $\mu $
from the equation \begin{equation}
\langle |\theta |^{2}\rangle \sim t^{\mu }\label{equ5.3}\end{equation}
 can be estimated as \begin{equation}
\mu =2\beta /\alpha \label{equ5.4}\end{equation}
 Expression (\ref{equ5.3}) should be considered with some reservations
since the second and higher moment may diverge. 

It was shown in \cite{Montroll84} that under special conditions the
exponent $\gamma $ for the trapping time asymptotic distribution
$\psi (t)\sim \tau ^{-\gamma }$ can be linked to fractal time dimension.
Moreover, $\gamma $ is related to the kinetic equation (\ref{eq:5.1})
as \cite{ZEN97}\begin{equation}
\beta =\gamma -1\, \, .\label{eq:5.16}\end{equation}

For the spatial distribution of particles, the simplest situation
occurs when the diffusion process is Gaussian for which $\alpha =2$.
When a hierarchical set of islands is present, $\alpha $ can be defined
through scaling properties of the island areas. In the considered
situation the random walk is more or less uniform in the regions where
trajectories are entangled near stable/unstable manifolds. That means
that we should expect that the value $\alpha \sim 2$ provides a good
estimate. Finally, we arrive to: \begin{equation}
\mu =2\beta /\alpha \sim \gamma -1\: .\label{eq:5.17}\end{equation}
 This last result is confirmed by all our observations for the different
vortex systems studied in \cite{KZ2000,LKZ01,LZ02}. We need to comment
that it is not worthwhile to try to obtain $\mu $ with a high accuracy
since a specific value of $\mu $ has no meaning due to multi-fractal
nature of transport \cite{Zaslavsky2000}. 

In fact, we can be even more precise, as the exponent $\gamma $ can
be estimated to $\gamma \approx 2.5$ leading to $\mu \approx 1.5$,
which is a good approximation of the different observed values given
the multifractality of the transport. We shall not reproduce here
the estimation of $\gamma $ (see for instance \cite{Zaslavsky2000,LKZ01}
for details), but the idea revolves around the presence of an island
of stability leading to ballistic or accelerator modes within the
island.

\section{Conclusion}

In this paper the dynamical and statistical properties of the passive
particle advection in flows generated by three, four and sixteen point
vortices has been reviewed. The goal of this work was to provide qualitative
insights on general transport properties of two-dimensional flows,
more specifically geophysical ones, imposed by the topology of the
phase space. The system of 16 vortices can be considered as a fairly
large system while the 3-vortex one is the minimal one with non fixed
distances between the vortices. Strong vortex-vortex correlation are
observed. These correlations manifest themselves the formation of
long lived pairs of vortices, triplets. Since these structures are
integrable, we can speculate that this form of stickiness occurs by
forming quasi-integrable subsets. The transport of passive tracers
is found to be anomalous and super-diffusive in all investigated situation
with a characteristic transport exponent $\mu \approx 1.5$. This
phenomenon is explained by the existence of coherent jets , which
are located in the sticky regions.

These jets and sticky region exhibit a complex fractal structures
in Figs \ref{cap:3vortfrac}, \ref{Figslowjet} and \ref{Fig16vjettime},
and are responsible for the anomalous behavior. Indeed hence the power
law behavior observed in Fig. \ref{Figescapetime4vortex} characterize
the exponent $\gamma $ for trapping time, and the similarly power-law
tails have been observed for Poincar\'{e} recurrence times distribution
in three vortex systems \cite{KZ2000,LKZ01}. In all situation we
observe an exceptional agreement with the $\gamma \approx \mu (2)-1$
relation resulting from the kinetic analysis performed in Sec. \ref{sec:Fractional-aspects-of}.
Hence since the notion of jet is quite general, we can say that the
anomalous diffusion finds its origin in the existence of jets typically
found around coherent structures (islands and cores). Moreover we
can access the transport properties of the global flow by simply gathering
escape time data from these coherent jets and using the equation (\ref{eq:5.17})
resulting from the fractional kinetic equation (\ref{eq:5.1}). 

We therefore emphasize that the present work by analyzing the role
played in transport by the different fractal structures involved in
our model-flows and by explicitly giving a typical value of the second
moment exponent as well as the trapping time exponent should be of
interest for the analysis of more realistic and complicated systems
involving many vortices and coherent structures such as geophysical
fluid dynamics.


\begin{thebibliography}{10}
\bibitem{Aref84}H.Aref, \emph{Stirring by chaotic advection}, J. Fluid Mech. \textbf{143},
1 (1984) 
\bibitem{Zaslav88}G. M. Zaslavsky, R. Z. Sagdeev, and A. A. Chernikov, Zhurn. Eksp.
Teor. Fiz. \textbf{94}, 102 (1988) (Soviet Physics, JETP \textbf{67},
270 (1988) 
\bibitem{Ottino89}J. Ottino, \textit{The kinematics of Mixing: Stretching, Chaos, and
Transport} (Cambrige U. P., Cambrige, 1989) 
\bibitem{Aref90}H.Aref, \emph{Chaotic advection of fluid particles}, Phil. Trans.
R. Soc. London \textbf{A 333}, 273 (1990) 
\bibitem{Ottino90}J. Ottino, Mixing, \emph{Chaotic advection and turbulence}, Ann. Rev.
Fluid Mech. \textbf{22}, 207 (1990) 
\bibitem{RomKedar90}V. Rom-Kedar, A. Leonard and S. Wiggins, \emph{An analytical study
of transport mixing and chaos in an unsteady vortical flow}, J. Fluid
Mech. \textbf{214}, 347 (1990) 
\bibitem{Zaslav91}G. M. Zaslavsky, R. Z. Sagdeev, D. A. Usikov, and A. A. Chernikov,
\emph{Weak Chaos and Quasiregular Patterns}, Cambridge Univ. Press,
(Cambridge 1991) 
\bibitem{Crisanti91}A. Crisanti, M. Falcioni, G. Paladin and A. Vulpiani, \emph{Lagrangian
Chaos: Transport, Mixing and Diffusion in Fluids}, La Rivista del
Nuovo Cimento, \textbf{14}, 1 (1991) 
\bibitem{Crisanti92}A. Crisanti, M. Falcioni, A. Provenzale, P. Tanga and A. Vulpiani,
\emph{Dynamics of passively advected impurities in simple two-dimensional
flow models}, Phys. Fluids A \textbf{4}, 1805 (1992) 
\bibitem{provenzale99}A. Provenzale, \emph{Transport by Coherent Barotropic Vortices}, Annu.
Rev. Fluid Mech. \textbf{31}, 55 (1999)
\bibitem{Majda99}A. J. Majda, J. P. Kramer, \emph{Simplified models for turbulent diffusion:
Theory, numerical modeling, and physical phenomena}, Phys. Rep. \textbf{314},
238 (1999) 
\bibitem{Weiss98}J. B. Weiss, A. Provenzale, J. C. McWilliams, \emph{Lagrangian dynamics
in high-dimensional point-vortex system}, Phys. Fluids \textbf{10},
1929 (1998) 
\bibitem{Tabeling98}P. Tabeling, A.E. Hansen, J. Paret, \emph{Forced and Decaying 2D turbulence:
Experimental Study}, in \textit{{}``Chaos, Kinetics and Nonlinear
Dynamics in Fluids and Plasma'', eds. Sadruddin Benkadda and George
Zaslavsky}, p. 145, (Springer 1998) 
\bibitem{Hansen98}A. E Hansen, D. Marteau , P. Tabeling, \emph{Two-dimensional turbulence
and dispersion in a freely decaying system}, Phys. Rev. E \textbf{58},
7261 (1998) 
\bibitem{Tabeling91}P. Tabeling, S. Burkhart, O. Cardoso, H. Willaime, \emph{Experimental-Study
of Freely Decaying 2-Dimensional Turbulence,} Phys. Rev. Lett. \textbf{67},
3772 (1991) 
\bibitem{Meleshko93}V.V. Meleshko, M.Yu. Konstantinov, \textit{Vortex Dynamics and Chaotic
Phenomena}, (World Scientific, Singapore, 1999)
\bibitem{Chernikov90}A. A. Chernikov, B. A. Petrovichev, A. V. Rogal'sky, R. Z. Sagdeev,
and G. M. Zaslavsky, \emph{Anomalous Transport of Streamlines Due
to their Chaos and their Spatial Topology}, Phys. Lett. A \textbf{144},
127 (1990) 
\bibitem{Solomon94}T.H. Solomon, E.R. Weeks, H.L. Swinney, \emph{Chaotic advection in
a two-dimensional flow: L\'{e}vy flights in and anomalous diffusion},
Physica D, \textbf{76}, 70 (1994) 
\bibitem{Weeks96}E.R. Weeks, J.S. Urbach, H.L. Swinney, \emph{Anomalous diffusion in
asymmetric random walks with a quasi-geostrophic flow example}, Physica
D, \textbf{97}, 219 (1996) 
\bibitem{Zaslavsky93}G.M. Zaslavsky, D. Stevens, H. Weitzner, \emph{Self-similar transport
in incomplete chaos}, Phys. Rev E \textbf{48}, 1683 (1993)
\bibitem{Kovalyov2000}S. Kovalyov, \emph{Phase space structure and anomalous diffusion in
a rotational fluid experiment}, Chaos \textbf{10}, 153 (2000) 
\bibitem{Benzi86}R. Benzi, G. Paladin, S. Patarnello, P. Santangelo and A. Vulpiani,
\textit{Intermittency and coherent structures in two-dimensional turbulence},
J. Phys A \textbf{19}, 3771 (1986) 
\bibitem{Benzi88}R. Benzi, S. Patarnello and P. Santangelo, \textit{Self-similar coherent
structures in two-dimensional decaying turbulence}, J. Phys A \textbf{21},
1221 (1988) 
\bibitem{Weiss87}J. B. Weiss, J. C. McWilliams, \textit{Temporal scaling behavior of
decaying two-dimensional turbulence}, Phys. Fluids A \textbf{5}, 608
(1992) 
\bibitem{McWilliams84}J. C. McWilliams, \textit{The emergence of isolated coherent vortices
in turbulent flow}, J. Fluid Mech. \textbf{146}, 21 (1984) 
\bibitem{McWilliams90}J. C. McWilliams, \textit{The vortices of two-dimensional turbulence},
J. Fluid Mech. \textbf{219}, 361 (1990) 
\bibitem{Elhmaidi93}D. Elhma\"{\i}di, A. Provenzale and A. Babiano, \textit{Elementary
topology of two-dimensional turbulence from a Lagrangian viewpoint
and single particle dispersion}, J. Fluid Mech. \textbf{257}, 533
(1993) 
\bibitem{Carnevale91}G. F. Carnevale, J. C. McWilliams, Y. Pomeau, J. B. Weiss and W. R.
Young, \textit{Evolution of Vortex Statistics in Two-Dimensional Turbulence},
Phys. Rev. Lett. \textbf{66}, 2735 (1991)
\bibitem{KZ98}L.~Kuznetsov and G.M.~Zaslavsky, \emph{Regular and Chaotic advection
in the flow field of a three-vortex system}, Phys. Rev E \textbf{58},
7330 (1998).
\bibitem{KZ2000}L. Kuznetsov and G. M. Zaslavsky, \emph{Passive particle transport
in three-vortex flow}. Phys. Rev. E. \textbf{61}, 3777 (2000).
\bibitem{LKZ2000}X. Leoncini, L. Kuznetsov and G. M. Zaslavsky, \emph{Motion of Three
Vortices near Collapse}, Phys. Fluids \textbf{12}, 1911 (2000)
\bibitem{LKZ01}X. Leoncini, L. Kuznetsov and G. M. Zaslavsky, \emph{Chaotic advection
near 3-vortex Collapse,} Phys. Rev.E, \textbf{63}, 036224 (2001)
\bibitem{Laforgia01}A. Laforgia, X. Leoncini, L. Kuznetsov and G. M. Zaslavsky, Eur. Phys.
J. B, \textbf{20}, 427 (2001)
\bibitem{LZ02}X. Leoncini and G. M. Zaslavsky, \emph{Jets, Stickiness, and anomalous
transport,} Phys. Rev.E, \textbf{65}, 046216 (2002)
\bibitem{Zabusky82}N. J. Zabusky, J. C. McWilliams, \textit{A modulated point-vortex
model for geostrophic, $\beta $-plane dynamics}, Phys. Fluids \textbf{25},
2175 (1982)
\bibitem{Vobseek97}P. W. C. Vobseek, J. H. G. M. van Geffen,, V. V. Meleshko, G. J. F.
van Heijst, \textit{Collapse interaction of finite-sized two-dimensional
vortices}, Phys. Fluids \textbf{9}, 3315 (1997)
\bibitem{VFuentes96}O. U. Velasco Fuentes, G. J. F. van Heijst, N. P. M. van Lipzig, \textit{Unsteady
behaviour of a topography-modulated tripole}, J. Fluid Mech. \textbf{307},
11 (1996)
\bibitem{Benzi92}R. Benzi, M. Colella, M. Briscolini, and P. Santangelo, \emph{A simple
point vortex model for two-dimensional decaying turbulence}, Phys.
Fluids A \textbf{4}, 1036 (1992)
\bibitem{Weiss99}J. B. Weiss, \emph{Punctuated Hamiltonian Models of Structured Turbulence},
in \textit{Semi-Analytic Methods for the Navier-Stokes Equations},
CRM Proc. Lecture Notes, \textbf{20}, 109, (1999)
\bibitem{Agullo01}O. Agullo and A. D. Verga Phys. Rev E \textbf{63,} 056304 (2001)
\bibitem{Aref79}H. Aref, \textit{Motion of three vortices}, Phys. Fluids \textbf{22},
393 (1979)
\bibitem{Aref83}H. Aref, \textit{Integrable, chaotic and turbulent vortex motion in
two-dimensional flows}, Ann. Rev. Fluid Mech. \textbf{15}, 345 (1983)
\bibitem{Novikov75}E. A. Novikov, \emph{Dynamics and statistics of a system of vortices},
Sov. Phys. JETP \textbf{41}, 937 (1975) 
\bibitem{Synge49}J. L. Synge, \textit{On the motion of three vortices}, Can. J. Math.
\textbf{1}, 257 (1949) 
\bibitem{Tavantzis88}J. Tavantzis and L. Ting, \textit{The dynamics of three vortices revisited},
Phys. Fluids \textbf{31}, 1392 (1988)
\bibitem{Novikov78}E. A. Novikov, Yu. B. Sedov, Stochastic properties of a four-vortex
system, Sov. Phys. JETP \textbf{48}, 440 (1978) 
\bibitem{ArefPomp82}H. Aref and N. Pomphrey, \emph{Integrable and chaotic motions of four
vortices: I. the case of identical vortices}, Proc. R. Soc. Lond.
A \textbf{380}, 359 (1982).
\bibitem{Ziglin80}S. L. Ziglin, Nonintegrability of a problem on the motion of four
point vortices, Sov. Math. Dokl. \textbf{21}, 296 (1980)
\bibitem{Babiano94}A. Babiano, G. Boffetta, A. Provenzale and A. Vulpiani, \emph{Chaotic
advection in point vortex models and two-dimensional turbulence},
Phys. Fluids \textbf{6}, 2465 (1994)
\bibitem{Boatto99}S. Boatto and R. T. Pierrehumbert, \emph{Dynamics of a passive tracer
in a velocity field of four identical point vortices.} J. Fuild Mech.
\emph{}\textbf{394}, 137 (1999).
\bibitem{Lamb45}H. Lamb, \textit{Hydrodynamics}, (6th ed. New York, Dover, 1945).
\bibitem{McLachlan92}R.I. McLachlan, P. Atela, \emph{The accuracy of symplectic integrators},
Nonlinearity \textbf{5}, 541 (1992).
\bibitem{Novikov79}E. A. Novikov, Yu. B. Sedov, \textit{Vortex collapse}, Sov. Phys.
JETP \textbf{22}, 297 (1979)
\bibitem{Kimura90}Y. Kimura, \textit{Parametric motion of complex-time singularity toward
real collapse}, Physica D \textbf{46}, 439 (1990)
\bibitem{Castiglione99}P. Castiglione, A. Mazzino, P. Mutatore-Ginanneschi, A. Vulpiani,
\emph{On Strong anomalous diffusion}, Physica D, \textbf{134}, 75
(1999)
\bibitem{Andersen00}K. H. Andersen, P. Castiglione, A. Massino, A. Vulpiani, \emph{Simple
stochastic models showing strong anomalous diffusion}, Eur. Phys.
J. B, \textbf{18}, 447 (2000)
\bibitem{Zaslavsky92}G.M. Zaslavsky, in \textit{{}``Topological Aspects of the Dynamics
of Fluids and Plasmas'', ed. H.K. Moffatt, et al}, p. 481, (Kluwer,
Dordrecht, 1992); Chaos \textbf{4}, 25 (1994); Physica D \textbf{76},
110 (1994) 
\bibitem{Chirikov79}B. V. Chirikov, Phys. Rep. \textbf{52}, 264 (1979) 
\bibitem{Rechester80}A.B. Rechester, R. White, Phys. Rev. Lett. \textbf{44}, 1586 (1980) 
\bibitem{dCN98}D. del-Castillo-Negrete, \emph{Asymmetric transport and non-Gaussian
statistics of passive scalars in vortices in shear}, Phys. Fluids
\textbf{10}, 576 (1998) 
\bibitem{Weitzner01}H. Weitzner and G. M. Zaslavsky, \emph{Directional fractional Kinetics},
Chaos \textbf{11}, 384 (2001)
\bibitem{Zaslav2000}G. M. Zaslavsky, \emph{Multifractional kinetics}, Physica A, \textbf{288},
431 (2000)
\bibitem{Zaslav2000_1}G. M. Zaslavsky, \emph{Fractional Kinetics of Hamiltonian Chaotic
Systems}, in {}`` Application of fractional Calculus in Physics'',
Ed. R. Hilfer (World Scientific, Singapore, 2000), p. 203 
\bibitem{Ferrari01}R. Ferrari, A. J. Manfroi, W. R. Young, \emph{Strong and weakly self-similar
diffusion}, Physica D, \textbf{154}, 111 (2001)
\bibitem{Afanasiev91}V. V. Afanasiev, R. Z. Sagdeev, and G. M. Zaslavsky, Chaos \textbf{1},
143 (1991)
\bibitem{Benkadda99}S. Benkadda, S. Kassibrakis, R. White and G. M. Zaslavsky, Phys. Rev.
E, \textbf{59}, 3761 (1999)
\bibitem{ZEN97}G. M. Zaslavsky, M. Edelman, B. A. Niyazov, \emph{Self-similarity,
renormalization, and phase space nonuniformity of Hamiltonian chaotic
dynamics}, Chaos \textbf{7}, 159 (1997)
\bibitem{romkedar99}V. Rom-Kedar and G. M. Zaslavsky, \emph{Islands of accelerator modes
and homoclinic tangles}, Chaos \textbf{9}, 697 (1999) 
\bibitem{Zaslavsky2000}G. M. Zaslavsky and M. Edelman, \emph{Hierarchical structures in the
phase space and fractional kinetics: I. Classical systems}, Chaos
\textbf{10}, 135 (2000)
\bibitem{Zasalvskypreprint}G. M. Zaslavsky, To appear in Physics Reports.
\bibitem{Hentschel84}H. G. E. Hentschel and I. Procaccia, Physica D \textbf{8}, 435 (1983);
P. Grassberger and I. Procaccia, \emph{ibid}, \textbf{13}, 34 (1984) 
\bibitem{Frish85}U. Frisch and G. Parisi, in \emph{Turbulence and Predictability of
Geophysical Flows and Climate Dynamics}, edited by M. Ghill, R. Benzi,
and G. Parisi (North-Holland, Amsterdam, 1985) 
\bibitem{Jensen85}M. H. Jensen, L. P. Kadanoff, A. Libshaber, I. Procaccia, and J. Stavans,
Phys. Rev. Lett. \textbf{55}, 439 (1985); T. C. Halsey, M. H. Jensen,
L. P. Kadanoff, A. Libshaber, I. Procaccia, and B. I. Schraiman, Phys.
Rev. A \textbf{33}, 1141 (1986) 
\bibitem{Montroll84}E. W. Montroll and M. F. Schlesinger, \emph{A wonderful World of Random
walk}, in {}``Studies in Statistical Mechanics'', edited by J. Lebowitz
and E. W. Montroll (North-Holland, Amsterdam, 1984), vol. 11, p. 1\end{thebibliography}
\end{document}